\documentclass[12pt]{article}

\advance \textheight by \topskip
\makeatletter
\@addtoreset{equation}{section}
 
\makeatother

\def\N{\mathbb N}
\def\Z{\mathbb Z}

\usepackage{amsmath,amssymb}

\begin{document}
\title{
{\bf Superconformal mechanics and nonlinear
supersymmetry}}

\author{{\sf Carlos Leiva${}^{a}$}\thanks{
E-mail: caleiva@lauca.usach.cl}
{\sf\ and Mikhail S. Plyushchay${}^{a,b}$}\thanks{
E-mail: mplyushc@lauca.usach.cl}
\\
{\small {\it ${}^a$Departamento de F\'{\i}sica,
Universidad de Santiago de Chile,
Casilla 307, Santiago 2, Chile}}\\
{\small {\it ${}^b$Institute for High Energy Physics,
Protvino, Russia}}}
\date{}






\maketitle


\begin{abstract}
We show that a simple change of the classical
boson-fermion coupling
constant,
$2\alpha \rightarrow 2\alpha n $, $n\in \N$, in the
superconformal mechanics model
gives rise to a radical
change of a symmetry:
the modified classical and quantum
systems are characterized by the nonlinear
superconformal symmetry.
It is generated by
the four bosonic integrals
which form the $so(1,2)\times u(1)$ subalgebra,
and  by the $2(n+1)$ fermionic integrals
constituting the two
spin-$\frac{n}{2}$
$so(1,2)$-representations
and anticommuting for the order
$n$ polynomials of the even generators.
We find that the modified
quantum system with an integer
value of the parameter $\alpha$
is described simultaneously
by the two nonlinear superconformal
symmetries of the orders relatively shifted in odd number.
For the original
quantum model with $|\alpha|=p$, $p\in \N$,
this means the presence
of the order $2p$ nonlinear superconformal
symmetry in addition to the
$osp(2|2)$ supersymmetry.
\end{abstract}

\newpage
\section{\protect\bigskip Introduction}

The conformal mechanics model of De Alfaro, Fubini and
Furlan
\cite{AFF} was introduced and studied as a $0+1$-dimensional
conformal
field theory. Its
supersymmetric analog was constructed
by Akulov and Pashnev \cite{AP}, and by Fubini and
Rabinovici
\cite{FR},  while the geometric nature of the original
and superextended versions of the model was established by
Ivanov, Krivonos and Leviant \cite{IKL}.
A recent revival of interest
in (super)conformal mechanics \cite{BH}-\cite{LP}
has been  triggered, in particular, by an
observation made in the
context of AdS/CFT correspondence conjecture \cite{AdS}
that the dynamics of a superparticle near the AdS
horizon of an extreme Reissner-Nordstr\"om black hole
is described by a superconformal mechanics
\cite{BH}.

There is, on the other hand,  an ongoing interest in
nonlinear generalization
of supersymmetric
quantum mechanics \cite{AIS}-\cite{AS}.
It is
caused mainly by the  exact
and quasi-exact solvability aspects of
the construction \cite{KP1,A1,KP2},
and by a recent observation of a reach algebraic structure
underlying nonlinear supersymmetry \cite{KP3}.
Such a generalization, however, has not touched yet
the superconformal symmetry.

The aim of the present paper is to
investigate the superconformal mechanics
in the light of nonlinear supersymmetry.
A priori it is not obvious whether
the corresponding generalization
of the superconformal mechanics exist
at least at the quantum level.
The reason is that unlike the standard
supersymmetric mechanics
possessing a linear superalgebraic structure \cite{Wit},
its nonlinear version is characterized by
the supercharges anticommuting for the
polynomials of the Hamiltonian
and of additional  even integrals if the
latter exist \cite{KP3}.
Due to a nonlinearity of the supercharges in the momenta,
generally a quantum anomaly prevents
a preservation of nonlinear supersymmetry
at the quantum level \cite{P1,KP1}.
We find that not only a nontrivial and unexpected
structure appears in the nonlinear generalization of
superconformal mechanics
already at the classical level:
the set of odd integrals is extended in
comparison with the original model \cite{AP,FR}
(cf. with nonlinear supersymmetry
of refs. \cite{KP1}-\cite{KP4},\cite{AS}),
but that the modified system is quantizable
in an anomaly free way.
Surprisingly, we find also that
at the special values
of the boson-fermion coupling constant,
in addition to the linear  $osp(2|2)$
Lie superalgebraic structure,
the original quantum
model \cite{AP,FR} itself
possesses the nonlinear superconformal symmetry
has not been known before.

The paper is organized as follows.
In Section 2 we review shortly the
$osp(2|2)$ Lie superalgebraic structure
of the classical version of the model
\cite{AP,FR}, and construct
the modified classical system
having the nonlinear superconformal symmetry.
We fix a general form of the corresponding
nonlinear Poisson bracket structure and then,
in Section 3, investigate
the quantum version of the modified system
and its nonlinear symmetry.
In Section 4 we find that there exist some
special values of the boson-fermion coupling
constant, for which the quantum system
is characterized simultaneously by
the two nonlinear superconformal symmetries
of different orders. In Section 5
we summarize the obtained results
and discuss some open problems to be interesting
for further investigation.


\section{Nonlinear classical superconformal mechanics}

In this section by an appropriate
change of the boson-fermion coupling constant,
we construct a classical modification of the
superconformal mechanics model
\cite{AP,FR} which is characterized by the odd integrals of
motion
to be polynomial in the momentum.
By explicit calculation we fix the
general Poisson structure
of the nonlinear superconformal symmetry
of the modified system.

\subsection{Linear superconformal mechanics}

To start, let us consider a free nonrelativistic particle
with spin
described by the Lagrangian
$$
L_0=\frac{1}{2}\dot{x}{}^2
-\frac{i}{2}\dot{\psi}_a\psi_a.
$$
Its  Hamiltonian has the same form as that
for the spinless particle,
$
H=\frac{1}{2}p^2.
$
In accordance with the equations of motion,
$\frac{dA}{dt}=\frac{\partial A}{\partial t}
+\{A,H\}$, generated via the
Poisson-Dirac brackets $\{x,p\}=1$,
$\{\psi_a,\psi_b\}=-i\delta_{ab}$,
$a,b=1,2$,
the
quantities
\begin{equation}
p,\quad
X=x-tp,\qquad
\psi_a,\quad
a=1,2,
\label{linint}
\end{equation}
form the set of integrals of motion
linear in phase space variables.
The integrals (\ref{linint}) generate the
space translations
($p$), the Galilei boosts
($X$), and the translations of odd variables
($\psi_a$).
The quadratic combinations of (\ref{linint}),
\begin{equation}
H=\frac{1}{2}p^2,\quad
D=\frac{1}{2}Xp=\frac{1}{2}xp-tH,\quad
K=\frac{1}{2}X^2=\frac{1}{2}x^2-2tD-t^2H,
\label{2even}
\end{equation}
\begin{equation}
\Sigma=\psi^+\psi^-=-i\psi_1\psi_2,
\label{sig}
\end{equation}
\begin{equation}
Q_a=p\psi_a,\qquad
S_a=X\psi_a=x\psi_a-tQ_a,
\label{qs}
\end{equation}
are also integrals of motion.
The even quadratic integrals
generate the translations in time ($H$),
the scale  (dilatation)
transformations ($D$), the special conformal
transformations ($K$), and the rotations of Grassmann
two-dimensional vector
$\psi_a$, or, equivalently, the U(1) phase
transformations
of the complex
Grassmann variables $\psi^\pm$  ($\Sigma$),
$
\psi^\pm=\frac{1}{\sqrt{2}}(\psi_1\pm i\psi_2),\quad
\{\psi^+,\psi^-\}=-i.
$
The odd integrals $Q_a$ and $S_a$
generate the
supertransformations
being the square roots from the time translations,
and from the special conformal transformations,
respectively.
The quadratic integrals form the classical
supeconformal algebra $osp(2|2)\cong su(1,1|1)$
(we write down only the nontrivial bracket relations):
$$
\{H,K\}=-2D,\quad
\{D,H\}=H,\quad
\{D,K\}=-K,
$$
$$
\{Q_a,Q_b\}=-2i\delta_{ab}H,\quad
\{S_a,S_b\}=-2i\delta_{ab}K,
$$
$$
\{S_a,Q_b\}=-2i\delta_{ab}D+i\epsilon_{ab}\Sigma,
$$
\begin{equation}
\{H,S_a\}=-Q_a,\quad
\{K,Q_a\}=S_a,
\label{scon0}
\end{equation}
$$
\{D,Q_a\}=\frac{1}{2}Q_a,\quad
\{D,S_a\}=-\frac{1}{2}S_a,
$$
$$
\{\Sigma,Q_a\}=\epsilon_{ab}Q_b,\quad
\{\Sigma,S_a\}=\epsilon_{ab}S_b.
$$
One could construct higher order even and odd
composite in  $p$, $X$ and $\psi^\pm$
integrals of motion.
In such a way, e.g.,
the super-Virasoro algebra
can be realized as a symmetry
on the phase space
of the free spin particle, see refs. \cite{Kum,CKZ}.

The case of the superconformal mechanics
given by the Hamiltonian
\begin{equation}
H=\frac{1}{2}\left(p^2+\frac{1}{x^2}
\alpha(\alpha+2i\psi_1\psi_2)\right)
\label{intsc}
\end{equation}
is less trivial: for such a system
none of the linear variables (\ref{linint})
is the integral of motion,
but there exist the analogs of the
quadratic integrals (\ref{2even})--(\ref{qs})
which form the same
superconformal algebra (\ref{scon0})\footnote{The
super-Virasoro
algebra may also be realized as a symmetry on the phase
space of the system (\ref{intsc}), see refs.
\cite{Kum,Marc}.}.
The corresponding generators of the
subalgebra $so(1,2)\times u(1)$
are the Hamiltonian (\ref{intsc}), and
\begin{equation}
D=
\frac{1}{2}xp-tH,\quad
K=
\frac{1}{2}x^2-2tD-t^2H,
\label{tdk}
\end{equation}
$$
\Sigma=\alpha-i\psi_1\psi_2.
$$
The odd generators of the $osp(2|2)$
take here the form
\begin{equation}
Q_a=p\psi_a+\frac{\alpha}{x}\epsilon_{ab}\psi_b,
\label{q1}
\end{equation}
\begin{equation}
S_a=(x-tp)\psi_a -\frac{\alpha}{x}t\epsilon_{ab}\psi_b.
\label{s1}
\end{equation}
All these integrals are polynomials in $p$
of order not higher than $2$,
and so, their quantum analogs are the local
differential operators.

\subsection{Nonlinear generalization}
Now, following ref. \cite{P1}
we shall show that a simple change of the
boson-fermion coupling constant,
$2\alpha\rightarrow 2\alpha n$, $n\in \N$,
i.e. the change of (\ref{intsc}) for the Hamiltonian
\begin{equation}
H_n=\frac{1}{2}
\left(p^2+\frac{1}{x^2}\alpha(\alpha+2in\psi_1
\psi_2)\right),
\label{hnclas}
\end{equation}
gives rise to a radical change of
the symmetry:
instead of the linear symmetry associated
with the Lie superalgebra
$osp(2|2)$, the system (\ref{hnclas}) is characterized
by the nonlinear superconformal symmetry
containing the same even subalgebra
$so(1,2)\times u(1)$,
but whose odd part includes  $2(n+1)$ generators
commuting (in the sense of classical Poisson brackets,
or, of quantum anticommutators)
for polynomials of even generators.

To reveal and analyze the superalgebraic structure of the
system (\ref{hnclas}), we define the even
oscillator-like variables
$$
z=\frac{\alpha}{x}+ip,\quad
\bar z=\frac{\alpha}{x}-ip,\quad
\{z,\bar z\}=2i\frac{\alpha}{x^2},
$$
in terms of which
the Hamiltonian (\ref{hnclas}) takes the form
$
H_n=\frac{1}{2}(z\bar z+in\{z,\bar z\}\psi^+\psi^-).
$
In addition  to the Hamiltonian and nilpotent charge
\begin{equation}
\Sigma=\psi^+\psi^-,
\label{sigm}
\end{equation}
the system (\ref{hnclas}) is characterized by the
even integrals $D_n$ and $K_n$ of the form
(\ref{tdk}) (with the Hamiltonian (\ref{hnclas})).
But, instead of the odd quantities (\ref{q1}),
here we have the integrals
\begin{equation}
S^+_{n,0}=z^n\psi^+,\quad S^-_{n,0}=
\bar z^n\psi^-,
\label{sn0}
\end{equation}
whereas the analogs of
the explicitly depending on $t$
integrals of motion (\ref{s1})  are
given by
\begin{equation}
S^+_{n,1}=\left(x+itz\right)z^{n-1}\psi^+,\quad
S^-_{n,1}=\left(x-it\bar z \right)\bar{z}{}^{n-1}
\psi^-.
\label{sn1}
\end{equation}
More exactly,
at $n=1$ the integrals (\ref{sn0}),
(\ref{sn1}) take the
form of the linear combinations
of the integrals (\ref{q1}), (\ref{s1}):
$S^+_{1,0}=\frac{i}{\sqrt{2}}(Q_1+iQ_2)$,
$S^+_{1,1}=\frac{1}{\sqrt{2}}(S_1+iS_2)$.
The  odd integrals (\ref{sn1})
are different from the integrals
(\ref{sn0}) in the factor
$$
\frac{x}{z}+it=\frac{\alpha-2iD_n}{2H_n}
$$
and in its complex conjugate, which is
the even (explicitly depending on time)
integral of motion.
Then,
one can extend  the odd integrals (\ref{sn0}),
(\ref{sn1})
by the set of odd integrals of the form
\begin{equation}
S_{n,l}^+=(x+itz)^lz^{n-l}\psi^+=\left(\frac{\alpha-2iD_n}{2
H_n}
\right)^lS^+_{n,0},
\quad S^-_{n,l}=(S^+_{n,l})^*,
\label{szt}
\end{equation}
where $l=2,\dots, n$.
The first equality from (\ref{szt}) means that all the
integrals
$S^\pm_{n,l}$, $l=0,1,...,n$, are polynomials
of the order $n$ in the momentum $p$.
Therefore, forgetting the anomaly
problem \cite{P1,KP1},
their quantum analogs
have to be the local differential operators of
the order $n$.
The second representation of $S^+_{n,l}$ from (\ref{szt})
will be  helpful  in finding the explicit form of
the superalgebra formed by all the set of
the even,
$H_n$, $D_n$, $K_n$, $\Sigma$,
and the odd, $S^\pm_{n,l}$, $l=0,\ldots, n$,
integrals, from which it will be clear why
the set of the integrals (\ref{sn0}), (\ref{sn1})
has to be extended
by the integrals (\ref{szt}) with $l=2,\ldots, n$.

First, we find that the bosonic part of the superalgebra
(\ref{scon0}) does not change,
\begin{equation}
\{H_n,K_n\}=-2D_n,\quad
\{D_n,H_n\}=H_n,\quad
\{D_n,K_n\}=-K_n.
\label{soclas}
\end{equation}
Then, taking into account the brackets
$
\{S^+_{n,0},S^-_{n,0}\}=-i(2H)^n,
$
and the relations
\begin{equation}
4(D_n^2-K_nH_n)-2n\alpha\Sigma=
-\alpha^2,
\label{casc}
\end{equation}
\begin{equation}
S^+_{n,0}S^-_{n,0}=(2H_n)^n\Sigma,
\label{ssh}
\end{equation}

one can find all other nontrivial brackets
of the integrals:
\begin{equation}
\{D_n,S^\pm_{n,l}\}=\left(\frac{n}{2}-l\right)
S^\pm_{n,l},\qquad
\{\Sigma,S^\pm_{n,l}\}=\mp iS^\pm_{n,l},
\label{dsig}
\end{equation}
\begin{equation}
\{H_n,S^\pm_{n,l}\}=\pm ilS^\pm_{n,l-1},\qquad
\{K_n,S^\pm_{n,l} \}=\pm i(n-l)S^\pm_{n,l+1},
\label{hkss}
\end{equation}

\begin{eqnarray}
\{S^+_{n,m},S^-_{n,l}\}=
&&-i(2H_n)^{n-m}(2K_n)^l(\alpha -2iD)^{m-l}
-i\Sigma(2H_n)^{n-m-1}(2K_n)^{l-1}\times
\nonumber\\
&&(\alpha -2iD_n)^{m-l}\left(n(m-l)(\alpha -2i D_n)+4\alpha
l(n-m)
\right),\quad
m\geq l.
\label{ss}
\end{eqnarray}
The brackets $\{S^+_{n,m},S^-_{n,l}\}$
for $m< l$ may be obtained
from (\ref{ss}) by complex conjugation
with application of  the relation
$\{B^*,A^*\}=-(\{A,B\})^*$
valid for arbitrary phase space functions $A$, $B$ of
fixed Grassmann parity.

The bracket relations mean that the
integrals of motion of the system (\ref{hnclas})
form the $\Z_2$-graded
nonlinear Poisson algebra:
the right hand side of Eq. (\ref{ss}) is polynomial in
the generators of the $so(1,2)\times u(1)$ subalgebra.

At first sight it seems that
at $l=0$ and at $m=n$
the nilpotent term (proportional to $\Sigma$)
on the right hand side of Eq. (\ref{ss})
contains a nonpolynomial factor.
However, the direct calculation with
using the relation (\ref{casc})
shows that in these two cases the brackets
(\ref{ss})
may be reduced to
the explicitly polynomial form,
\begin{eqnarray}
&\{S^+_{n,m},S^-_{n,0}\}=-i(2H_n)^{n-m}
(\alpha-2iD_n)^{m-1}
\left((\alpha -2iD_n)
+nm\Sigma\right),&\label{ss1}\\
&
\{S^+_{n,n},S^-_{n,l}\}=
-i(2K_n)^l(\alpha-2iD_n)^{n-l-1}\left(
(\alpha-2iD_n)+n(n-l)\Sigma\right).
&
\label{ss2}
\end{eqnarray}

According to the relations (\ref{hkss}), the Hamiltonian
acts on the set  $S^\pm_{n,l}$
as a decreasing in index $l$ generator,
``annihilating" the lowest in $l$ odd integrals
$S^\pm_{n,0}$ which do not depend,
unlike the $S^\pm_{n,l}$, $1\leq l\leq n$,
explicitly on time.
On the other hand,
$K_n$ acts as an increasing in $l$
generator annihilating the highest in $l$ odd
generators $S^\pm_{n,n}$.
Having also in mind the first relation from
(\ref{dsig}),
we conclude that the sets of odd generators
$S^+_{n,l}$ and $S^-_{n,l}$ are transformed with respect
to the action of the $so(1,2)$ generators as the
two  $(n+1)$-dimensional multiplets
(spin-$\frac{n}{2}$ ``states'').
Therefore, the subsequent action (via the Poisson brackets)
of the conformal boost generator $K_n$
on the not depending explicitly on time
integrals $S^\pm_{n,0}$ generate all the
finite set of the odd integrals $S^\pm_{n,l}$
terminating the series on $l=n$.

In accordance with
relations (\ref{casc}), (\ref{ssh}),
the classical nonlinear
superconformal algebra
(\ref{soclas}), (\ref{dsig}), (\ref{hkss}),
(\ref{ss})
has the Casimir elements
$D_n^2-K_nH_n-\frac{1}{2}\alpha n \Sigma$ and
$S^+_{n,0}S^-_{n,0}-(2H_n)^n\Sigma$
taking here the values $-\frac{1}{4}\alpha^2$
and $0$, respectively.
Analogously, the even combinations
$
S^+_{n,l}S^-_{n,l}-(2H_n)^{n-l}(2K)_n^l\Sigma,
$
$l=1,\ldots,n$,
are also classical central elements of the
algebra taking a zero value.

\section{Quantum case}
As we already mentioned,
the specific feature  of
nonlinear
supersymmetry,
unlike the linear $n=1$ case,
consists in appearance of
the quantum anomaly problem that destroys
the conservation of the odd generators at the quantum level
\cite{P1,KP1}.
However, there exist some classes of
superpotentials for which the problem of the quantum
anomaly may be resolved \cite{KP1}-\cite{KP2}.
Below we show that the system (\ref{hnclas})
belongs to such a class.

\subsection{Quantum integrals of motion}

Let us choose the quantum analog of the  Hamiltonian
(\ref{hnclas})  in the form
\begin{eqnarray}
&H_n=\frac{1}{2}\left(-\frac{d^2}{dx^2}+
\frac{1}{x^2}(a_n +b_n\sigma_3)\right),&\label{hq}
\\
&
a_n=\alpha_n^2+\frac{1}{4}(n^2-1),\quad
b_n=-n\alpha_n,\quad
\alpha_n=\alpha-\frac{1}{2}(n-1),
&
\label{ab}
\end{eqnarray}
where $\Delta_1=a_n-\alpha^2$ and $\Delta_2=b_n+n\alpha$ are
the
quantum corrections \cite {KP1}
(here we put $\hbar=1$) which, as will be shown,
guarantee the conservation of the
nonlinear superconformal symmetry.

First we note that
the quantum analogs of the generators
(\ref{tdk}), (\ref{sigm})
of the classical bosonic symmetry $so(1,2)\times u(1)$,
\begin{equation}
D_n=\frac{1}{4}[x,p]_{{}_+}-tH_n,\quad
K_n=\frac{1}{2}x^2-2tD_n-t^2H_n,\quad
\Sigma=\frac{1}{2}[\psi^+,\psi^-]=\frac{1}{2}\sigma_3,
\label{qbos}
\end{equation}
are the integrals of motion of the quantum system
(\ref{hq}). Any of them satisfies the equation of the form
\begin{equation}
i\frac{\partial I_n}{\partial t}-[H_n,I_n]=0.
\label{in}
\end{equation}
The quantum system (\ref{hq}) admits also the set
of odd integrals being the quantum analogs of the
classical integrals $S^\pm_{n,l}$,
$l=0,1,\ldots,n$.
These are given by the relations
\begin{eqnarray}
&S^+_{n,0}={\cal P}_{\alpha, n}\psi^+,&
\label{s0q}\\
&S^+_{n,l}=(x+it{\cal D}_{\alpha -n+1})(x+it{\cal D}_{\alpha
-n+2})...
(x+it{\cal D}_{\alpha -n+l+1})
{\cal P}_{\alpha, n-l}\psi^+,\quad
l=1,\ldots, n,&
\label{slq}
\end{eqnarray}
where
$$
{\cal P}_{\alpha, n} ={\cal D}_{\alpha -n+1}{\cal D}_{\alpha
-n+2}...{\cal D}_{\alpha -1}
{\cal D}_\alpha,\qquad
{\cal D}_\gamma=\frac{d}{dx}+\frac{\gamma}{x},\qquad
\psi^\pm=\frac{1}{2}(\sigma_1\pm i\sigma_2).
$$
To prove the conservation of the odd operators
(\ref{s0q}), (\ref{slq}),
we use the relation
\begin{equation}
{\cal D}_{\alpha -k}(x+it
{\cal D}_{\alpha -k+1})=(x+it
{\cal D}_{\alpha -k}){\cal D}_{\alpha -k+1}
\label{dd}
\end{equation}
to  get the identities
\begin{equation}
\frac{\partial S^+_{n,l}}{\partial t}=ilS^+_{n,l-1}
\label{sl1}
\end{equation}
and
\begin{equation}
S^+_{n,l}=itS^+_{n,l-1}+xS^+_{n-1,l-1},
\label{snl}
\end{equation}
which are valid for $l=1,\ldots, n$.
Next, we note that
$[H_1,S^+_{1,0}]=0$
and suppose that
 $[H_{n-1},S^+_{n-1,0}]=0$
 is valid for $n\geq 1$.
Then, using the relation
\begin{equation}
H_n=H_{n-1}+(n-\alpha-1)\frac{1}{x^2}\Pi_+,
\label{hn}
\end{equation}
where
$
\Pi_+=\frac{1}{2}(1+\sigma_3),
$
$\Pi_+\psi^+=\psi^+$,
$\psi^+\Pi_+=0$,
we find that
$[H_n,S^+_{n,0}]=0$. Therefore, by induction,
we conclude that $S^+_{n,0}$ commutes with
corresponding $H_n$
for any $n\in\N$.
A simple
calculation shows that
$S^+_{n,1}$ satisfies the equation of the form (\ref{in}),
and so, is the quantum integral of motion
which, like $D_n$
and $K_n$, contains
an explicit dependence on time.
Then, assuming that $S^+_{n,l-1}$ satisfies equation
of the form (\ref{in}),
and using the relations (\ref{snl}) and (\ref{hn}),
we find that $S^+_{n,l}$ also satisfies the same equation,
and by induction conclude that
$S^+_{n,l}$, $l=1,2,\ldots, n$,
are the integrals of motion.

\subsection{Quantum nonlinear superconformal algebra}

The quantum integrals of motion satisfy
the nonlinear superconformal algebra
similar to the classical one given by Eqs.
(\ref{soclas}), (\ref{dsig}), (\ref{hkss}),
(\ref{ss}).
Before the discussion of its structure,
we note that it is sufficient
to find the form of any (anti)-commutation
relation for $t=0$ while its validity for arbitrary $t$
is a consequence of the unitary evolution,
$I_n(t)=U_n(t)I_n(0)U_n^{-1}(t)$, $U_n(t)=\exp (-itH_n)$,
of any integral of motion $I_n$ obeying the equation
(\ref{in}).

The bosonic operators
(\ref{qbos})  of the  $so(2,1)\times u(1)$
subalgebra satisfy the relations (only a nontrivial part)
\begin{equation}
[H_n,K_n]=-2iD_n,\quad
[D_n,H_n]=iH_n,\quad
[D_n,K_n]=-iK_n.
\label{qso}
\end{equation}
Using the explicit form of the even integrals,
we find that the quantum analog of the Casimir element
(\ref{casc}) takes here the following value:
\begin{equation}
2(H_nK_n+K_nH_n)-4D_n^2+2n\alpha_n \Sigma  =
\alpha_n^2+\frac{1}{4}n^2-1,
\label{qcas}
\end{equation}
where $\alpha_n$ is given by Eq. (\ref{ab}).

The nontrivial commutators between even and odd
integrals are
\begin{equation}
[\Sigma,S^\pm_{n,l}]=\pm S^\pm_{n,l},\quad
[D_n,S^\pm_{n,l}]=i\left(\frac{n}{2}-l\right)
S^\pm_{n,l},
\label{sds}
\end{equation}
\begin{equation}
[H_n,S^\pm_{n,l}]=\mp lS^\pm_{n,l-1},\quad
[K_n,S^\pm_{n,l}]=\mp (n-l)S^\pm_{n,l+1}.
\label{hks}
\end{equation}
The commutation relations of $H_n$ with the
odd generators
follow immediately from
the equations (\ref{in}) and (\ref{sl1}).
The commutation relations (\ref{hks})
between $K_n$ and $S^\pm_{n,l}$ for
$l=n$ and $l=n-1$ can easily be proved by a direct
calculation. Then, assuming that (\ref{hks}) is valid for
$l=k+1$, and using the relation
$$
S^+_{n,l}=\frac{1}{x}{\cal D}_{\alpha -n}S^+_{n,l+1}
$$
and its Hermitian conjugate, we prove the validity
of the formula for $l=k$, that by induction
finishes the proof of the second commutation relations
from (\ref{hks}) for arbitrary
$l$, $l=0,\ldots, n$.

{}From the dimensionality of the integrals
(given by the commutator with $D_n$)
it follows that the anticommutators between odd integrals
coincide up to the ordering and quantum corrections
to the parameter $\alpha$ (see Eq. (\ref{qcas}))
with the classical relations (\ref{ss}).
For the simplest nonlinear case $n=2$ these are
$$
[S^+_{2,0},S^-_{2,0}]_{{}_+}=(2H_2)^2,\quad
[S^+_{2,2},S^-_{2,2}]_{{}_+}=(2K_2)^2,\quad
[S^+_{2,1},S^-_{2,1}]_{{}_+}=(2D_2)^2+\alpha_2^2,
$$
$$
[S^+_{2,2},S^-_{2,0}]_{{}_+}=\left(2iD_2-\alpha_2
\right)
\left(2iD_2-\alpha_2-4\Sigma\right),\quad
[S^-_{2,2},S^+_{2,0}]_{{}_+}=([S^+_{2,2},S^-_{2,0}]_{{}_+})^
\dagger,
$$
$$
[S^+_{2,1},S^-_{2,0}]_{{}_+}=
-2i(D_2H_2+H_2D_2)+2H_2\left(\alpha_2+2\Sigma\right),
\quad
[S^-_{2,1},S^+_{2,0}]_{{}_+}=([S^+_{2,1},S^-_{2,0}]_{{}_+})^
\dagger,
$$
$$
[S^+_{2,2},S^-_{2,1}]_{{}_+}=
-2i(D_2K_2+K_2D_2)+2K_2\left(\alpha_2+2\Sigma\right),\quad
[S^-_{2,2},S^+_{2,1}]_{{}_+}=([S^+_{2,2},S^-_{2,1}]_{{}_+})^
\dagger,
$$
where $\alpha_2$ is given by Eq. (\ref{ab})
with $n=2$.
Comparing with classical relations, we see that
the anticommutation relations are obtained from the
corresponding
Poisson brackets (\ref{ss}), (\ref{ss1}), (\ref{ss2}) by
symmetrization of the product of the
noncommuting $so(1,2)$ generators
and by the quantum shift
$\alpha\rightarrow \alpha-\frac{1}{2}$.

In general case of supersymmetry of the order $n$,
one can easily find the following anticommutation relations:
\begin{equation}
[S^+_{n,n},S^-_{n,n}]_{{}_+}=(2K_n)^n,
\label{ssk}
\end{equation}
\begin{equation}
[S^+_{n,n},S^-_{n,n-1}]_{{}_+}=-i
(D_n (2K_n)^{n-1}+(2K_n)^{n-1}D_n)+(2K_n)^{n-1}(\alpha_n+n
\Sigma).
\label{ssdk}
\end{equation}
According to relation (\ref{ssdk}),
the quantum analog of the
classical Poisson bracket relations
is obtained by the same quantization prescription as in
the simplest nonlinear case $n=2$: by symmetrizing the
product of noncommuting operators and by changing $\alpha$
for
$\alpha_n$.
However, comparing the quantum relation
\begin{eqnarray}
[S^+_{n,n-1},S^-_{n,n-1}]_{{}_+}=&&\frac{1}{2}
(2H_n(2K)^{n-1}+(2K_n)^{n-1}2H_n)+(n-1)(n-2)(2K_n)^{n-2}
\nonumber\\
&&+4(n-1)(2K)^{n-2}\alpha_n\Sigma,
\label{n-1}
\end{eqnarray}
with the corresponding classical relation
(\ref{ss}) taken for $m=l=n-1$,
one concludes that due to the presence of the
second term,
the same quantization
prescription (a simple symmetrization of  the
noncommuting factors accompanied by the
shift of the parameter $\alpha$)
does not work in general case
(note, that the additional term in (\ref{n-1})
vanishes
for the simplest nonlinear case $n=2$).
Similar complications appear also
in writing the explicit structure of other anticommutators
of the odd generators for the general case
of the nonlinear superconformal symmetry
of the order $n$.
On the other hand, the anticommutator of not depending
explicitly on time
odd integrals  is, like Eq. (\ref{ssk}), the direct quantum
analog of the
corresponding
classical Poisson bracket relations:
\begin{equation}
[S^+_{n,0},S^-_{n,0}]_{{}_+}=(2H_n)^n.
\label{s0s0}
\end{equation}
To conclude the discussion of the structure
of the quantum nonlinear superconformal algebra,
we prove the relation (\ref{s0s0}).
The two terms on the left hand side of Eq.
(\ref{s0s0}) may be written in the form
\begin{equation}
l.h.s.= (-1)^n{\cal P}_{\alpha,n}{\cal P}_{-\alpha +n-1,n}
\Pi_+
+(-1)^n{\cal P}_{-\alpha+n-1,n}{\cal P}_{\alpha,n}\Pi_-,
\label{lhs}
\end{equation}
where in correspondence with the notation used above
$\Pi_\pm=\frac{1}{2}(1\pm \sigma_3)$, $\Pi_\pm^2=\Pi_\pm$,
$\Pi_+\Pi_-=0$,
$\Pi_++\Pi_-=1$.
Subsequently using the relation
$$
{\cal D}_\gamma{\cal D}_{-\gamma}=
{\cal D}_{-(\gamma-1)}{\cal D}_{\gamma-1},
$$
Eq. (\ref{lhs}) may be represented as
$$
l.h.s.=(-{\cal D}_{-(\alpha-n)}{\cal D}_{\alpha -n})^n\Pi_+
+(-{\cal D}_{-\alpha}{\cal D}_{\alpha})^n\Pi_-.
$$
The Hamiltonian (\ref{hq}) has an equivalent representation
\begin{equation}
H_n=\frac{1}{2}\left(-\frac{d^2}{dx^2}+\frac{j_n(j_n+1)}{x^2
}\right),
\qquad j_n=\alpha -n\Pi_+,
\label{hj}
\end{equation}
from which we get the equalities
$2H_n\Pi_+=-{\cal D}_{-(\alpha-n)}{\cal D}_{\alpha -n}\Pi_+$
and
$2H_n\Pi_-=-{\cal D}_{-\alpha}{\cal D}_{\alpha}\Pi_-$,
and so, reduce Eq. (\ref{lhs}) to the right hand side of
Eq. (\ref{s0s0}).

\section{Two related nonlinear superconformal symmetries}

Relation (\ref{hn}) means that for the special value of the
parameter, $\alpha=n-1$, the equality $H_n=H_{n-1}$
takes place.
This means that the corresponding system
may be characterized simultaneously by the nonlinear
superconformal symmetries of the orders $n$ and $n-1$.
Let us investigate the nature of such systems
and the associated supersymmetries.
To do this, we first investigate the question
what is the most general case of the quantum systems
(\ref{hq}) possessing such a property.
If the system can be characterized by the supersymmetries
of different orders $n$ and $n'$,
its Hamiltonian $H_n$ has to admit
the alternative representation in the form
of the Hamiltonian $H_{n'}$ corresponding
in general case to a different value of the
parameter $\alpha$.
So, let us require that
\begin{equation}
H_n(\alpha)=H_{n'}(\alpha ').
\label{hh}
\end{equation}
Having in mind that Eq. (\ref{hh})
can be satisfied in two ways,
$H^\pm_n(\alpha)=H^\pm_{n '}(\alpha ')$,
or
$H^\pm_n(\alpha)=H^\mp_{n '}(\alpha ')$,
where the index $+$ ($-$) corresponds to the $+1$
($-1$) eigensubspace of $\sigma_3$,
we find the following possible cases:

\begin{eqnarray}
&\alpha=\alpha '=\frac{1}{2}(n+n '-1),&
\label{a1}\\
&\alpha=-(\alpha '+1)=\frac{1}{2}(n- n' -1).&
\label{a2}
\end{eqnarray}
The Hamiltonian in these two cases is given by
\begin{equation}
H=\frac{1}{2}\left(-\frac{d^2}{dx^2}+\frac{(n\pm n')^2-1}
{4x^2}\mp \frac{nn'}{x^2}\Pi_+\right),
\label{hami1}
\end{equation}
where the upper and lower signs correspond to the cases
(\ref{a1}) and (\ref{a2}).
Two other cases are given by
the relations
\begin{eqnarray}
&\alpha -n=-(\alpha '+1)=\frac{1}{2}(n+n '+1),&\label{a3}\\
&\alpha-n=\alpha '=\frac{1}{2}(n- n' +1),&
\label{a4}
\end{eqnarray}
\begin{equation}
H=\frac{1}{2}\left(-\frac{d^2}{dx^2}+\frac{(3n
+2\pm n')^2-1}
{4x^2}- \frac{n(2n+2\pm n')}{x^2}\Pi_+\right).
\label{hami2}
\end{equation}
Comparing (\ref{hami1}) with (\ref{hami2}),
we see that the second Hamiltonian can be obtained
from the first one via the formal change
$n'\rightarrow n'\mp 2(n+1)$.
Note that the case (\ref{hn})
is a particular case of (\ref{a1})
with $n'=n-1$, and that in all the cases
(\ref{a1}), (\ref{a2}), (\ref{a3}), (\ref{a4})
the model parameter
$\alpha$
takes integer or half-integer
values.

Let us consider some particular cases.
When $n'=0$, the last term in Eq. (\ref{hami1})
turns into zero, and the Hamiltonian takes a pure bosonic
form
\begin{equation}
H=\frac{1}{2}\left(-\frac{d^2}{dx^2}+\frac{n^2-1}{4x^2}
\right).
\label{hbos}
\end{equation}
This corresponds to a trivial supersymmetry
with odd integrals of motion $\psi^+$ and $\psi^-$.
This is in correspondence with Eqs. (\ref{hq}),
(\ref{ab}) according to which
the Hamiltonian
takes a pure bosonic form ($b_n=0$)
only for $\alpha=\frac{1}{2}(n-1)$,
i.e. when $\alpha$ takes an integer or a half-integer value.
On the other hand, this system can be characterized
by the nonlinear superconformal supersymmetry of
the order $n$.
Since $\psi^\pm$ are the odd integrals of motion,
the corresponding even operators
appearing as factors in odd integrals
(\ref{s0q}), (\ref{slq})
are the integrals of motion.
Let us clarify their nature.
For this,
subsequently using the factorization relations
\begin{equation}
-\frac{d^2}{dx^2}+\frac{n^2-1}{4x^2}=-
{\cal D}_{\frac{1}{2}(n+1)}
{\cal D}_{-\frac{1}{2}(n+1)}=
-{\cal D}_{-\frac{1}{2}(n-1)}
{\cal D}_{\frac{1}{2}(n-1)}
\label{factor}
\end{equation}
we find that in the case of $n=2p$
the odd integral $S^+_{2,0}$ is
equal to $(-2H)^p\psi^+$.
The same happens with
other odd integrals of motion:
they are reduced to the products of
the trivial
odd integrals $\psi^\pm$ and of the
even integrals $H$, $D$ and $K$.
Therefore, in the case $n=2p$ the order $n$ supersymmetry
of the system (\ref{hbos}) is trivial.

The pure bosonic system
(\ref{hbos}) with odd $n=2p+1$  is more interesting.
In this case the integral of motion ${\cal P}_{p,2p+1}$
presenting as a factor in odd integral
$S^+_{2p+1,0}$ is a a local differential operator
of order $2p+1$ which  effectively
is the
$p+\frac{1}{2}$ degree of the Hamiltonian.
For instance, for $n=5$ this follows from the following
chain of relations due to a subsequent application
of Eq. (\ref{factor}):
\begin{eqnarray}
&({\cal P}_{2,5})^2={\cal D}_{-2}{\cal D}_{-1}{\cal D}_{0}
{\cal D}_{1}({\cal D}_{2}{\cal D}_{-2}){\cal D}_{-1}
{\cal D}_{0}{\cal D}_{1}{\cal D}_{2}=
{\cal D}_{-2}{\cal D}_{-1}{\cal D}_{0}
{\cal D}_{1}({\cal D}_{-1}{\cal D}_{1}){\cal D}_{-1}
{\cal D}_{0}{\cal D}_{1}{\cal D}_{2}=&\nonumber\\
&{\cal D}_{-2}{\cal D}_{-1}{\cal D}_{0}
{\cal D}_{0}^2{\cal D}_{0}^2
{\cal D}_{0}{\cal D}_{1}{\cal D}_{2}=
{\cal D}_{-2}{\cal D}_{-1}({\cal D}_{1}
{\cal D}_{-1})^3{\cal D}_{1}{\cal D}_{2}=&\nonumber\\
&{\cal D}_{-2}({\cal D}_{2}{\cal D}_{-2})^4
{\cal D}_{2}=({\cal D}_{3}{\cal D}_{-3})^5=
(-2H)^5&
\end{eqnarray}
Analogously, other odd operators,
$S^\pm_{n,l}$, $l=1,\ldots,n$, supply us with
the even polynomial differential operators being
effectively the half-integer degrees of the  operators
$H^{n-m}K^{m}$ (in this symbolic writing we neglect the
quantum ordering problem).

One can observe that
in general cases of (\ref{a1}),
(\ref{a2}), (\ref{a3}) and (\ref{a4})
with $n'\neq 0$, $n>n'$,
the following relation takes place:
\begin{equation}
S^+_{n,0}={\cal A}_{n,n'}S^+_{n',0}.
\label{cala}
\end{equation}
When $n-n'=2p$, $p=1,2,\ldots$,
this relation reduces to
$S^+_{n,0}=(-2H)^pS^+_{n',0}$,
and higher nonlinear supersymmetry of order $n$
is a trivial derivative of the lower supersymmetry.
The simplest example of such a case is given
by Eqs. (\ref{a1}) and (\ref{hami1})
with $n=3$, $n'=1$, $\alpha=\alpha'=\frac{3}{2}$,
for which, with taking into account Eq.
(\ref{factor}), we have the relation
$2H\psi^+=-{\cal D}_{-\frac{1}{2}}
{\cal D}_{\frac{1}{2}}\psi^+$,
and, as a consequence,
$S^+_{3,0}=-2HS^+_{1,0}$.
On the other hand,
the cases (\ref{a1}),
(\ref{a2}), (\ref{a3}) and (\ref{a4})
with $n'\neq 0$, $n-n'=2p+1$
give the systems with the both nonlinear supersymmetries
of the orders $n'$ and $n=n'+2p+1$ to be nontrivial.
As an illustration, one can consider the
system with $n'=n-1=\alpha$.
Such a system is given by the Hamiltonian
\begin{equation}
H=\frac{1}{2}\left(-\frac{d^2}{dx^2}+\frac{n(n-1)}{x^2}\Pi_-
\right)
\label{hpi-}
\end{equation}
and is characterized by the two superconformal
symmetries of the orders $n$ and $n-1$.
Note that the upper component $H^+$ of the Hamiltonian
(\ref{hpi-}) corresponds to the free particle system.
The anticommutators of the supercharges
corresponding to the
$n$ and $n-1$ supersymmetries
produce even nontrivial integrals of motion.
For the simplest case of $n=2$,
there is the relation
$S^+_{2,0}={\cal D}_0S^+_{1,0}$.
Since $2H\psi^+=-{\cal D}_0^2\psi^+$,
we see that here the role of the additional factor
in $S^+_{2,0}$ is reduced effectively to multiplication of
the supercharge $S^+_{1,0}$ by the square root from the
Hamiltonian.
The anticommutator of the supercharges $S^+_{2,0}$ and
$S^-_{1,0}$ produces the even integral of motion,
$$
[S^+_{2,0},S^-_{1,0}]_{{}_+}=
-\left({\cal D}_0^3-\frac{3}{x^2}{\cal D}_{-1}\Pi_-\right)
\equiv {\cal R},
$$
and the direct calculation
shows that the third order local differential operator
${\cal R}$ has a nature of the $3/2$ degree of the
Hamiltonian:
${\cal R}^2=(-2H)^3$.
Analogously, other anticommutators
of the $S^\pm_{2,l}$, $l=0,1,2$,
with $S^\pm_{1,l'}$, $l'=0,1$,
produce even integrals being effectively
half-integer degrees of the
third order polynomials
of the $so(1,2)\times u(1)$ generators.

The considered example of the system (\ref{hpi-})
with $n'=1$ and $n=2$ belongs to the general class of the
systems given by the Hamiltonian of the form
\begin{equation}
H=\frac{1}{2}\left(-\frac{d^2}{dx^2}+\frac{p}
{x^2}(p- \sigma_3)\right),\quad
p\in \N.
\label{hamik}
\end{equation}
The Hamiltonian (\ref{hamik}) is a particular case
of the system (\ref{a1}), (\ref{hami1})
with $n'=1$ and $n=2p$,
and is the direct quantum analog
of the classical system (\ref{intsc})
with $\alpha=p$.
In addition
to the usual superconformal symmetry
of the order $n'=1$, the quantum system
(\ref{hamik}) possesses the nonlinear
superconformal symmetry of the order $n=2p$,
the odd generators of which produce,
via the anticommutators with the
fermionic $n'=1$ superconformal symmetry generators,
the additional nontrivial
bosonic integrals of motion having a
nature of the half-integer degrees
of the
odd order polynomials
of the
$so(1,2)\times u(1)$ generators.

\section{Discussion and outlook}

We have showed that a simple change
of the boson-fermion coupling
constant
$2\alpha\rightarrow 2\alpha n$,
$n\in \N$, in  the
classical superconformal mechanics
model (\ref{intsc})
gives rise to a radical change of the symmetry.
Instead of the linear Lie superalgebraic $osp(2|2)$
structure associated with the model
(\ref{intsc}), the modified system
(\ref{hnclas}) is characterized by
the nonlinear superconformal symmetry of
the order $n$.
The latter is generated
by the four even integrals forming the
$so(1,2)\times u(1)$ Lie subalgebra,
and by the two complex conjugate
sets of the $(n+1)$ odd integrals forming the
spin-$\frac{n}{2}$
representations of the $so(1,2)$ subalgebra and
anticommuting (in the Poisson bracket sense)
for the polynomials of the order $n$ of even generators.

We have demonstrated that there exists
the anomaly free
quantization
of the modified system preserving
all the set of the integrals
and the nonlinear superalgebraic structure.
On the other hand,
due to the quantum corrections a new,
in comparison with the classical case,
phenomenon arises at the quantum level:
when the model parameter $\alpha$ takes an integer value,
the quantum system is characterized
simultaneously by the two nonlinear
superconformal symmetries of the orders relatively
shifted in odd number
(see, however, ref. \cite{KP1} where the classical
analog of such a situation was discussed
and refs. \cite{ACIN,AS}, where a similar phenomenon
was discussed at the quantum level for
some systems with $N=1$ nonlinear supersymmetry).
The anticommutators of
the generators from these two different fermionic sets
produce new nontrivial bosonic local integrals of
motion being effectively
the half-integer degrees of the
odd order polynomials
of the
$so(1,2)\times u(1)$
generators.
In particular case, when the Hamiltonian has a purely
bosonic form (\ref{hbos}),
one of the corresponding supersymmetries
of the order $n'=0$ is trivial
 (the corresponding odd integrals
are reduced to the  fermionic operators $\psi^\pm$),
and fermionic generators of another supersymmetry
of the order $n=2p+1$ are
reduced to the  compositions of the
conserved fermionic operators $\psi^\pm$
and of additional nontrivial bosonic integrals.
(When $n=2p$, the even factors in the corresponding
fermionic generators of
the order $n=2p$ supersymmetry are
reduced to the products of the $so(1,2)$ generators,
i.e. the additional nonlinear superconformal symmetry
is trivialized).
Another interesting particular case of the
systems with the two supersymmetries is given
by the Hamiltonian (\ref{hamik}).
This quantum family,
being, on the one hand, a quantum analog of the
original system
(\ref{intsc}) with $\alpha=p$, $p\in \N$,
possesses a usual $n'=1$
superconformal symmetry and, on the other hand,
is characterized simultaneously  by the
nonlinear superconformal symmetry of
the order $n=2p$.

Recently it has been shown \cite{Kum,CKZ,CCM,Iva2}
that the ``relativistic" generalization
of the superconformal mechanics model
which  governs the motion of a superparticle
near the AdS  horizon of an extreme
Reissner-Nordstr\"om black hole \cite{BH}
is canonically equivalent to the original model
\cite{AP,FR}.
Therefore, for the appropriate values
of the parameters of the ``relativistic"
superconformal mechanics system of ref.
\cite{BH}
(when the corresponding boson-fermion coupling constant
takes an even integer value), at the quantum level
in addition to the linear $osp(2|2)$
Lie superalgebraic symmetry
it possesses a nontrivial nonlinear superconformal
symmetry of the even order.

To conclude, let us
list some problems which deserve a
further attention.

The superconformal mechanics model
describes the relative motion of the two-particle
Calogero system.
Therefore, it
would be interesting to investigate
whether the $N$-particle
supersymmetric Calogero system
\cite{FM,GT,Wyl,bgk}
admits a generalization for the case
of the nonlinear supeconformal symmetry.
Another interesting problem is a geometric
realization of the described nonlinear superconformal
symmetry on the appropriate superspace.
This, in particular, could help in clarifying
the question of
the superfield formulation for the model
(\ref{hnclas}).
The obtained results may also be
developed in application
to other supersymmetric
systems possessing superconformal symmetries,
e.g., to the fermion-monopole system \cite{HL,PMF}.
We hope to present the results of the analysis
of these problems elsewhere.

\vskip 0.5cm
{\bf Acknowledgements}
\vskip 5mm
We  thank D. Sorokin for a useful comment.
The work has been supported in part by
FONDECYT-Chile (grant 1010073) (M.P.),
and by CONICYT-Chile (C.L.).

\vskip 0.2cm

{\bf Note added.} In the subsequent paper \cite{anpl}, the
origin of the quantum corrections to the integrals
of motion  of the model (\ref{hnclas}) is clarified, and the
complete form of the quantum nonlinear superconformal
algebra  of the arbitrary order $n$ is fixed.

\end{document}